# Swelling and shrinking kinetics of a lamellar gel phase

David J Fairhurst[1,2a)], Mark E Baker[3], Neil Shaw[3], Stefan U Egelhaaf[2,4]

[1]*Department of Physics, Nottingham Trent University, Nottingham, NG11 8NS, UK*
[2]*SUPA, School of Physics, The University of Edinburgh, Edinburgh, EH9 3JZ, UK*
[3]*Unilever Research, Port Sunlight Laboratory, Bebington, Wirral L63 3JW, UK*
[4]*Condensed Matter Physics Laboratory, Heinrich-Heine-University, 40225 Düsseldorf, Germany*

We investigate the swelling and shrinking of $L_\beta$ lamellar gel phases composed of surfactant and fatty alcohol after contact with aqueous poly(ethylene-glycol) solutions. The height change $\Delta h(t)$ is diffusion-like with a swelling coefficient, $S$: $\Delta h = S\sqrt{t}$. On increasing polymer concentration we observe sequentially slower swelling, absence of swelling, and finally shrinking of the lamellar phase. This behavior is summarized in a non-equilibrium diagram and the composition dependence of $S$ quantitatively described by a generic model. We find a diffusion coefficient, the only free parameter, consistent with previous measurements.

In everyday life and many industrial processes, materials swell by absorption of solvent, e.g. washing powder,[1] foodstuffs,[2] diapers,[3] eyeballs[4] and clay[5]. Conversely, if the solvent flow is reversed materials shrink, as for a hypertonic cell with a lower solute concentration than its environment. Model systems are often preferred for study: swelling rates of $L_\alpha$ surfactant lamellar phases are observed to change when the chemical potential difference between lamellar phase and contacting solution is varied through polymer addition;[6] artificial liposomes can be swollen or shrunk using glycerol solutions;[7] hard sphere colloidal suspensions shrink when contacted with high concentration polymer solutions.[8] Here we quantitatively investigate the swelling and shrinking behavior of a complex surfactant system, namely an $L_\beta$ lamellar gel phase, as used in cosmetics and pharmaceuticals. The volume change is initiated by contact with aqueous polymer solution. Our observations are in quantitative agreement with a generic model, which we expect to be applicable to a large variety of situations.

The lamellar phase is prepared following an industrial procedure.[9,10] It consists of a cationic quaternary surfactant, behenyl trimethyl ammonium chloride (BTAC) and a fatty alcohol, 1-octadecanol, at a molar ratio of 1:3 in water with different total surfactant concentrations $c_s$. Electron and light microscopy reveal a disordered system: numerous stacks of bilayers form an open structure with small pockets of water and excess fatty alcohol.[10] While the sample is prepared at elevated temperature $T$, the experiments are performed at $T = 25°C$, which is below the chain melting temperature (about 78°C). Approximately 0.5 g of this $L_\beta$ lamellar gel phase is pipetted into a 2 cm$^3$ cylindrical glass cell and centrifuged for 1 min at 2500 rpm to ensure that the entire highly viscous sample is at the bottom of the cell with a smooth upper surface. Within a range of lamellar masses (0.45-1.25 g) no systematic change in behavior was observed within the ~12% experimental uncertainty. The experiment is started by adding 1.5 g of water or polymer solution on top of the lamellar phase. The use of polymer (poly(ethylene glycol), PEG-10000 with molar mass from 8500 to 11500 g/mol and an average radius of gyration of about 3 nm) allows us to vary the difference in chemical potential, i.e. osmotic pressure or water concentration, between the two phases.

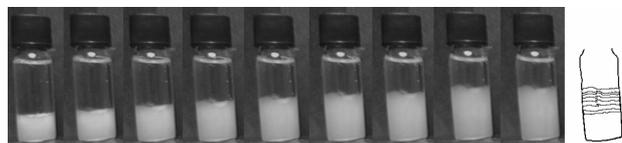

FIG 1. Swelling of lamellar phase with an initial total surfactant concentration $c_s$ = 6% w/w. To illustrate the square-root growth behavior, images at times $t$ = 0, 1, 4, 9, 16, 25, 36, 49 and 64 h (left to right) are shown. The right-most image is the digitized version used for the analysis.

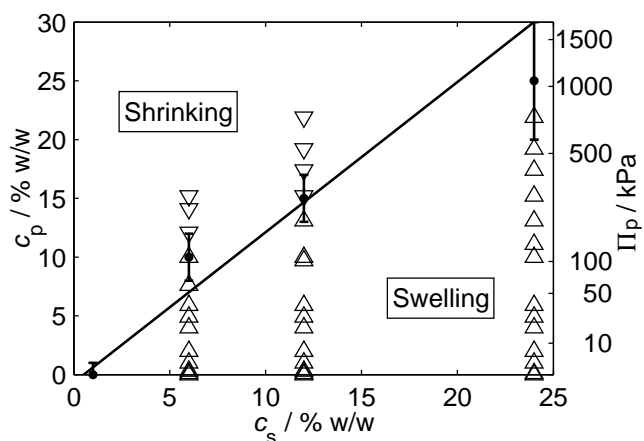

FIG 2. Non-equilibrium diagram in the initial total surfactant concentration ($c_s$)-polymer concentration ($c_p$) plane indicating the compositions of samples that swell (upwards triangles) or shrink (downwards triangles). Solid circles indicate where no change in volume is expected, based on data in Fig.4. The solid line represents the fitted boundary $c'_p(c_s)$ according to Eq. 1. The right hand axis indicates the corresponding osmotic pressure of the polymer solution, $\Pi_p(c_p)$.

After contact with solvent, the lamellar phase remains as one contiguous mass, but changes its volume on a time scale of hours (Fig. 1). Depending on the polymer concentration $c_p$ of the contacting solution, qualitatively different behavior is observed: the lamellar phase swells for $c_p$ below a specific concentration $c'_p(c_s)$ (Fig. 2, upward triangles); in contrast, for $c_p > c'_p(c_s)$ the lamellar phase shrinks, (downward triangles). Based on the experimentally determined $c'_p(c_s)$ values and the additional observation that a sample with $c_s$ =

---

[a] Author to whom correspondence should be addressed; electronic mail: david.fairhurst@ntu.ac.uk

1% does not show detectable swelling when contacted with water, we obtain by a fit

$$c'_p(c_s) = 1.28 c_s - 0.65\% \text{ w/w} \qquad (1)$$

where we assumed a linear dependence between $c'_p$ and $c_s$. Along this boundary (Fig. 2, solid line), the sample neither swells nor shrinks, which suggests that the osmotic pressures of the surfactant phase $\Pi_s(c'_s)$ and contacting polymer solution $\Pi_p(c'_p)$ are balanced; $\Pi_s(c'_s) = \Pi_p(c'_p)$. Using the known osmotic pressure $\Pi_p(c_p)$ of our polymer solution[11] (Fig. 2, right hand axis) and Eq. 1, we can calculate $\Pi_s(c_s)$, e.g $\Pi_s(6\%) = 52$ kPa. With no literature data for $\Pi_s(c_s)$, we calculate the osmotic pressure of an ideal gas of chloride counter-ions (from BTAC, $c_{BTAC} = 49$ mM): $\Pi_{Cl}(6\%) = RTc_{BTAC} = 121$ kPa with universal gas constant R. Given this crude approximation, the agreement with $\Pi_s(6\%)$ is encouraging. Poisson-Boltzmann theory provides a more accurate calculation[12] but requires a precise knowledge of the surface charge density and bilayer spacing. Nevertheless, if not only $\Pi_p(c_p)$ but also $\Pi_s(c_s)$ is known *a priori* or can be measured or calculated, one could, based on $\Pi_s(c'_s) = \Pi_p(c'_p)$, compute the location of the boundary $c'_p(c_s)$ and thus predict the behavior of any sample.

Upon contact, the concentration of the lamellar phase at the interface with the polymer solution must jump to the relevant equilibrium value $c'_s(c_p)$. Assuming polymer does not move into the lamellar phase, i.e. $c_p$ is constant, the value of $c'_s(c_p)$ is found graphically by moving horizontally on Fig. 2 from the initial point $(c_s, c_p)$ to the boundary $(c'_s, c_p)$ and mathematically by inverting Eq. 1 to get $c'_s(c_p)$. The abrupt change in surfactant concentration at the interface is unstable and decays through counter-diffusion of water and surfactant. For $c_s > c'_s$, the surfactant concentration has to decrease to $c'_s$ so water will enter the lamellar phase causing it to swell. Conversely, for $c'_s > c_s$ the interface is at a higher surfactant concentration than the bulk lamellar phase and water will diffuse out, resulting in a shrinking lamellar phase.

We now investigate the swelling and shrinking kinetics quantitatively. The change in volume, or for our sample geometry the change in sample height $\Delta h$, is followed (Fig. 1, right-most image). For all samples, $\Delta h$ as a function of time since contact, $t$, can be described by

$$\Delta h(t) = S\sqrt{t} \qquad (2)$$

with swelling coefficient $S$ (Fig. 3, solid lines) which is positive for samples that swell and negative for those that shrink. This form of growth is common in many systems, including the swelling of $L_\alpha$ lamellar phases[6,13,14] the swelling of polymer gels[4] and capillary flow.[15] Fits to each $\Delta h(t)$ data set provide the dependence of $S$ on $c_p$ and $c_s$ (Fig. 4), with $S(c_s, c_p)$.[16] To predict the $c_s$- and $c_p$-dependence of $S(c_s, c_p)$ we use its relation to the diffusion coefficient $D$. Numerical methods were used to extract $D$ from the motion of interfaces in a swelling $L_\alpha$ lamellar system.[14] Here we adapt an analytical solution to this Stefan (moving boundary) problem, previously used for the precipitation of a solid phase from a supersaturated liquid,[17] and the growth of a colloidal crystal[18]. This requires two assumptions: both phases are semi-infinite and thus the concentrations $c_s$ and $c_p$ are fixed far from the interface (only at late times do we observe sub-diffusive growth caused by the limited extent of the two phases); polymer diffusion is slow into the lamellar phase, i.e. into the gaps between bilayers, but fast within the contacting (bulk) polymer solution relative to the movement of the interface, thus $c_p$ is constant in both time and space. Then $S(c_s,c_p)$ can be related to $D$ by:[18]

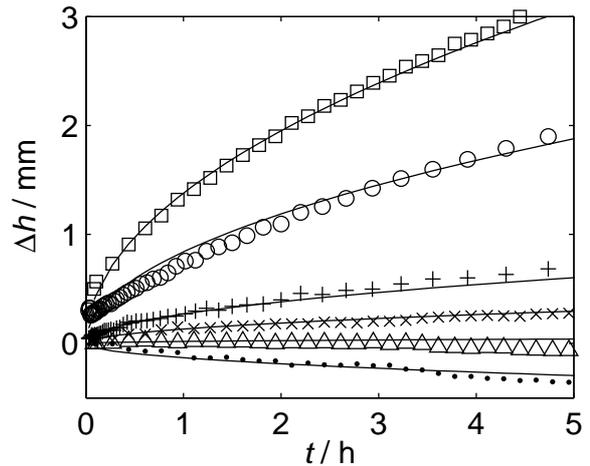

FIG 3. Change in sample height $\Delta h$ as a function of time since contact, $t$, for lamellar phase (initial total surfactant concentration $c_s$ = 6% w/w) contacted with polymer solutions with concentration $c_p$ = 0, 0.3, 4.9, 7.6, 10.0 and 15.2 w/w, top to bottom. The solid lines are fits for the swelling coefficients $S$: data up to $t = 40$ h were used, but only early times are shown.

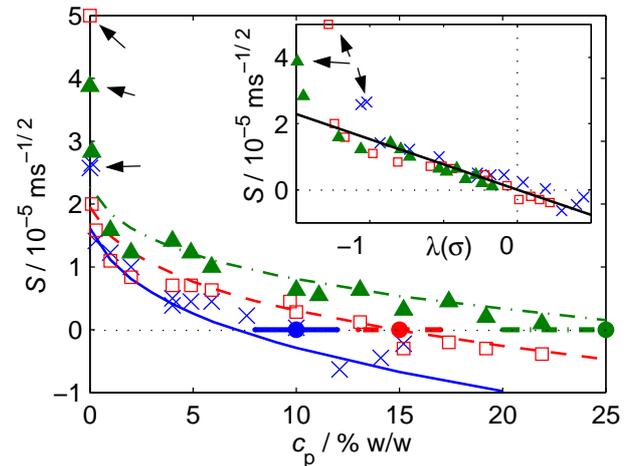

FIG 4. Measured swelling coefficients $S$ as a function of polymer concentration $c_p$. The lines are fits with $D = 5.8\times10^{-11}$ m$^2$s$^{-1}$ as the only free parameter. Colors (online), symbols and lines are: initial total surfactant concentration $c_s$ = 6%: blue, crosses, solid line; 12%: red, hollow squares, dashed line; 24%: green, filled triangles, dash-dotted line. Solid circles show estimate of where each data-set cross the x axis, with thick horizontal lines indicating uncertainties. The inset shows the same values of $S(c_p)$ plotted against the velocity coefficient $\lambda(\sigma) = f^{-1}(\sigma)$, where $\sigma$ is the relative supersaturation (Eqs. 3 and 4). The best fit line has a negative slope of $(1.525 \pm 0.070) \times 10^{-5}$ ms$^{-1/2} = 2\sqrt{D}$. Samples contacted with pure water are indicated by arrows.

$$S(c_s, c_p) = -2\lambda(c_s, c_p)\sqrt{D(c_s)} . \qquad (3)$$

The velocity coefficient $\lambda(c_s,c_p)$[19] is related to the relative supersaturation $\sigma(c_s,c_p) = \left(c'_s(c_p) - c_s\right)/c'_s(c_p)$ by

$$\sigma(c_s, c_p) = \sqrt{\pi}\lambda e^{\lambda^2} \text{erfc}(\lambda) . \qquad (4)$$

For each sample we use the inverted Eq. 1 to calculate $\sigma(c_s,c_p)$ and then find $\lambda(c_s,c_p)$ numerically from Eq. 4. Eq. 3 suggests that on a graph of $S(c_s,c_p)$ versus $\lambda(c_s,c_p)$ the experimental points for all samples should collapse onto a single straight line through the origin with slope $-2\sqrt{D}$. This is indeed observed (Fig. 4, inset), with significant deviations only for the samples contacted with pure water ($c_p = 0$, arrows), which are ignored in the following. If we fit individual $D(c_s)$ values for each $c_s$, we obtain similar values. A $D$ independent of $c_s$ is often assumed *a priori* as it reduces mathematical complexities and seems not overly restrictive,[20] so we fit a single value to all data sets irrespective of $c_s$ and obtain $D = (5.8\pm0.5)\times10^{-11}$ m$^2$s$^{-1}$. The agreement between fit and experimental data (Fig. 4, main plot and inset) is remarkable given the simplicity of the model, the complexity of the lamellar gel phase and the presence of only one adjustable parameter, $D$. The value of $D$ can be compared with values for other systems obtained with different, more involved techniques: for AOT solutions forming various phases, $0.2 – 9\times10^{-11}$ m$^2$s$^{-1}$;[13] for an aqueous $C_{12}E_6$ solution forming a lamellar $L_\alpha$ phase, $8.5 – 12\times10^{-11}$ m$^2$s$^{-1}$;[21] for $C_{12}E_5$ solutions with concentrations between 5% and 50% w/w, $5 – 20\times10^{-11}$ m$^2$s$^{-1}$.[22] Thus, our value of $D$ is not unreasonable.

In conclusion, we contacted $L_\beta$ lamellar gel phases consisting of BTAC and fatty alcohol with aqueous polymer solutions. Depending on whether the polymer concentration $c_p$ was below or above a specific concentration $c'_p(c_s)$, the volume of the lamellar phase either increased or decreased. This was summarized in a non-equilibrium diagram (Fig. 2) and rationalized in terms of the osmotic pressure difference between the lamellar phase and the polymer solution. Swelling and shrinking kinetics were both described by a square-root time dependence with a swelling coefficient $S(c_s,c_p)$ (Fig. 3). Based on an established theoretical framework,[18] we collapsed $S(c_s,c_p)$ for *all* $c_s$ and $c_p$ onto a single line (Fig. 4, inset). The slope of this line is related to the diffusion coefficient $D$, independent of $c_s$ and $c_p$ and the only free parameter in the model. A fit resulted in $D = (5.8\pm0.5)\times10^{-11}$ m$^2$s$^{-1}$, which is consistent with values for other systems.

The theoretical approach is independent of the specific details of the system. It describes the process after contacting any two phases which can be considered semi-infinite and where the solute concentration in the contacting solution can be assumed constant (here $c_p$): First, based on the osmotic pressures of the initial phases (here $\Pi_s$ and $\Pi_p$) or, if they are unknown, on the experimentally determined boundary dividing swelling and shrinking behavior, the final equilibrium composition $c'_p(c_s)$ can be determined (line in Fig. 2, Eq. 1). Then the supersaturation $\sigma(c_s,c_p)$ and velocity coefficient $\lambda(c_s,c_p)$ can be determined (Eq. 4). Together with the diffusion coefficient $D$, which is available for some systems[13,21,22] or can be estimated, the swelling coefficient $S(c_s,c_p)$ can be calculated (Eq. 3), which determines the time-dependence of the sample height $\Delta h(t)$ (Eq. 2) and thus the volume change as a function of the initial concentrations, $c_s$ and $c_p$. We maintained the subscripts 's' and 'p', although the two phases do not need to be surfactant and polymer phases. It is merely required that the two phases exchange only solvent, as through a semi-permeable membrane. Due to this generic nature, we believe the theoretical approach to be potentially applicable to many other contact, dilution, dissolution and swelling situations.

We thank Unilever plc and the Engineering and Physical Sciences Research Council (EPSRC Grant GR/S10377, Edinburgh Soft Matter and Statistical Physics Programme Grant) for support, M E Cates, J Leng and A Wagner for very helpful discussions and P B Warren and R M L Evans for critical comments on the manuscript.